\documentclass[10pt,a4paper,x11names]{article}
\usepackage[utf8]{inputenc}
\usepackage{amsmath}
\usepackage{amsfonts}
\usepackage{amssymb}
\usepackage[top=2cm, bottom=2cm, left=2cm, right=2cm, includehead, includefoot, heightrounded]{geometry}
\usepackage{pagecolor}
\usepackage{fancyhdr}
\usepackage[style=ieee,sorting=none,url=true,doi=true,backend=bibtex]{biblatex}
\usepackage{tikz}
\usetikzlibrary{arrows.meta}
\usetikzlibrary{automata, shapes}
\usetikzlibrary{decorations.markings}
\usetikzlibrary{calc}
\usepackage{pdflscape}
\usepackage{makecell}
\usepackage{hhline}
\usepackage{tabularx}
\usepackage{booktabs}
\usepackage{titlesec}
\usepackage{ltablex}
\definecolor{lightblue}{HTML}{F7FAFC}
\definecolor{blue}{HTML}{3182ce}
\definecolor{darkblue}{HTML}{2a4365}
\pagecolor{lightblue}

\usepackage[defaultfam,tabular,lining]{montserrat}
\usepackage[T1]{fontenc}

\titlespacing*{\paragraph}{0pt}{0.5ex plus 1ex minus .2ex}{1em}

\author{Q Misell\\University of Aberdeen}
\title{Hiding in \texttt{text/plain} sight: Security defences of Tor Onion Services}

\begin{document}

\maketitle{}

\begin{abstract}
Tor Onion Services are a way to host websites and other internet services anonymously. Onion Services are often used to bypass internet censorship and provide information services to users in oppressive regimes. This paper presents an analysis of the security defences deployed on these Onion Services. Onion Services tend to have better security policy than sites on the clear web. However they lag behind in the deployment of HTTPS, a key defence to ensuring the security of users of such services.
\end{abstract}

\section{Introduction}
Tor (alternatively The Onion Router) is a decentralised network of hosts allowing communication across the internet in which neither the sender nor the receiver know anything about each other. It achieves this by encapsulating data in layers of encryption (the onion concept), and transiting data over at least 3 nodes. Each node is aware of its neighbours, such as the data's source or final destination, but no node is ever aware of both the source and destination of the data\cite{dingledine2004tor}.

The Tor network supports so-called Onion Services. These act in most ways like normal websites, but provide much stronger cryptographic guarantees about privacy and data integrity than traditional internet websites. When an Onion Service is configured correctly it is near impossible to track who connects to it, and where the server hosting the site is\cite{onion-services-overview}.

Onion Services provide a vital way to circumvent internet censorship in repressive regimes and get accurate news into countries that lack freedom of the press\cite{bbg-iac}. Many newsrooms have a presence on the Onion Services network, including the BBC, Deutsche Welle, The Guardian, and ProPublica\cite{bbc-tor-mirror,dw-tor-mirror,guardian-tor-mirror,propublica-tor-mirror}. The anonymity of Onion Services also lends itself well to less ethical activities\cite{guitton2013review}, however this should not draw away from its benefits to society.

Given the danger users can put themselves in by accessing content banned in their country\cite{kravchenko2019russian}, its is imperative that these users stay as anonymous as possible. It is also important that operators of Onion Services remain as anonymous as possible, less they are targeted by regimes unfriendly to them\cite{rahimi2003cyberdissent}. This research aims to analyse various security and privacy methods employed by Onion Services, their prevalence across the Onion Service network, and categorise which types of sites tend to employ which methods.

\section{Research question}
This paper aims to analyse the security defences deployed on Onion Services, and compare them to the state of security defences on sites on the clear web. The hypothesis considered in this paper are as follows:

\begin{enumerate}
\item Onion Services are as likely to use HTTPS as other sites.
\item Single Onion Services are as likely to use HTTPS as other Hidden Services.
\item Onion Services use the same versions of TLS as other sites.
\item Onion Services use the certificate key lengths as other sites.
\item Onion Services are as likely to use Extended Validation certificates as other sites.
\item Onion Services are as likely to use Strict Transport Security as other sites.
\item Onion Services are as likely to use a Content Security Policy as other sites.
\item Onion Services are as likely to use a Permissions Policy as other sites.
\item Onion Services are as likely to use X-Frame-Options as other sites.
\item Onion Services are as likely to use X-Content-Type-Options as other sites.
\item Onion Services are as likely to use a Referrer Policy as other sites.
\end{enumerate}

\subsection{Scope}
Onion Services can host a multitude of services capable of running over TCP, not just websites using HTTP(S)\cite{rend-spec-v3}. In order to limit the scope of work only HTTP services on the Tor network will be analysed. The class of security and privacy problems than can apply to internet services other than HTTP are distinct and out of scope for this paper.

\section{Related work}
\subsection{OnionScan}
OnionScan\footnote{\url{https://onionscan.org}} was a project that enumerated all Onion Services hosting websites and scanned for security vulnerabilities. Since the depreciation of version 2 Onion Services in 2020\cite{torv2-deprecation} its no longer possible to enumerate all of the Onion Services on the Tor network, due to the use of a blinded public key for the \texttt{.onion} address generation\cite{rend-spec-v3}. 

The analysis completed by OnionScan provides an insight into what data would be useful to collect about Onion Services to analyse their security, and methods to locate Onion Services to analyse\cite{onionscan-what}.

\subsection{Qualys SSL Labs}
Qualys SSL Labs\footnote{\url{https://www.ssllabs.com/ssltest/}} is a tool for analysing the correctness of a site's implementation of TLS for HTTPS. Whilst there may appear to be no security benefit of using HTTPS over Tor, as connections to Onion Services are already end-to-end encrypted\cite{rend-spec-v3}, this is not the case\cite{real-world-onion}. Of particular note is that the Tor tunnel may terminate before the HTTP reaches the final web server (as in the case of sites using the Enterprise Onion Toolkit\footnote{\url{https://github.com/alecmuffett/eotk}}), and HTTPS provides security guarantees about this final step in accessing the site. A user may also not connect directly to the Tor network when accessing a Onion Service but instead use a Tor proxy somewhere on their local network via SOCKS5\cite{rfc1928}; HTTPS secures this part of the connection as well. Qualys SSL labs tests a website for various possible misconfiguration of its TLS setup that could impact privacy or security of the site or its users\cite{ristic2012ssl}.

Browsers also behave very differently for HTTP and HTTPS sites. Features such as microphone and webcam access, WebUSB, and geo-location (to name but a few) are only available on HTTPS enabled sites\cite{securecontext}. While the Tor Browser may make patches to enable such features non-HTTPS Onion Services, the Tor Project team is a lot smaller than the teams behind the likes of Mozilla Firefox and Google Chrome. Using the codepath that's been developed by the vastly more mature team with stricter code review flows will likely result in better security overall.

Whilst SSL Labs is incapable of scanning Onion Services, their reports on the state of TLS on the clear web provide insight into what misconfigurations should be investigated in Onion Services.

\subsection{Security Headers}
Servers can send various HTTP headers back to the client to enforce security policies on their site, to aid in preventing man-in-the-middle, code injection, etc. TLS protects the contents of a data exchange, but does not entirely influence the security policy applied by a web browser\cite{buchanan2018analysis}. Security Headers\footnote{\url{https://securityheaders.com}} is a tool developed by Scott Helme that grades sites on their use of various security policy headers.

Security Headers is incapable of scanning Onion Services, but provides insight into useful metrics to analyse, and datasets from it can be used as a comparator for security between Onion Services and sites on the clear web.

\subsection{Discovering Onion Services}
Onion Services are - by their very nature - hidden. This makes collecting a list of Onion Services to analyse a somewhat difficult prospect. It would be impossible to know of every Onion Service in existence. Onion Service owners do however have to publicise their Onion Services URL somewhere, or nobody would ever connect to it. Owners often publish their Onion Service URLs on sites such as The Hidden Wiki\footnote{\url{https://thehiddenwiki.org}, further context \url{https://web.archive.org/web/20150628213012/https://www.deepdotweb.com/2014/11/15/the-hidden-wiki-seized/}}, or submit them to the Onion Search Engine\footnote{\url{https://onionsearchengine.com}}\cite{matic2017active}. These collections can be used to gather a list of sites to analyse.

Certificate Transparency logs can also be used a source to discover Onion Services\cite{scheitle2018rise}. For any website to obtain a publicly trusted TLS certificate the certificate must be submitted to a public and searchable log of certificates.

\section{Methodology}
This steps required to collect data for this research can be broken down into the following broad stages:
\begin{enumerate}
\item collation of a list of Onion Services
\item collection of data about each Onion Service discovered
\item collection of the same data about clear web sites
\end{enumerate}

\subsection{Collecting Onion Services}
A web crawler was developed that searches through Onion Services for links to other Onion Services, akin  to how Googlebot\cite{googlebot} - and other search engines - crawl and discover content on the public internet. Given the slow speed of connections over the Tor network\cite{measuring-tor}, and the long time it takes to initiate connections to new Onion Service; the collection was run in a highly parallelised manner with 50 Tor clients and 50 crawlers balancing load between the Tor clients. This was all coordinated by message passing over RabbitMQ. The source code for this crawler is available on GitHub\footnote{\url{https://github.com/TheEnbyperor/shadow-weaver}}.

Certificate Transparency logs were also searched also be searched for any certificates issued to Onion Services. Certificate Transparency is a log consisting of every publicly trusted X.509 certificate ever issued\cite{laurie2014certificate}, and thus included every Onion Service with a publicly trusted certificate. To complete this search Sectigo's crt.sh\footnote{\url{https://crt.sh}} was used, a database of every certificate ever logged in any widely trusted Certificate Transparency log.

\subsubsection{Limitations}
This methodology of discovering Onion Services does have its limitations. Firstly the web crawling approach only discovers sites that people actively publicise on other sites. This is true of the clear web too, however Onion Services wish to stay entirely hidden, as would be expected from sites using such facilities as Tor. This likely had an effect on the type of sites discovered, and the number.

Discovering sites via Certificate Transparency likely resulted in an over-count of the prevalence of HTTPS on Onion Services. The dataset used for analysis will contain every single Onion Service with HTTPS, but not contain every single one without.

\subsection{Collection of data about Onion Services}
The following data was collected about each Onion Service in the discovered dataset.
\begin{enumerate}
\item which ports are open
\begin{enumerate}
\item HTTP (port 80)
\item HTTPS (port 443)
\item SSH (port 22)
\item FTP (port 21)
\item SMTP (port port 25)
\end{enumerate}
\item if HTTPS is offered
\begin{enumerate}
\item which versions of TLS are supported
\item supported TLS cipher suites
\item downgrade attack prevention\cite{rfc7507}
\item support for TLS compression
\item support for TLS 1.3 early data
\item is the service vulnerable to Heartbleed
\item is the service vulnerable to CCS injection
\item is the service vulnerable to Client Renegotiation DOS
\item is the service vulnerable to ROBOT
\item is the X.509 certificate presented an extended validation certificate
\item is the X.509 certificate presented a publicly trusted certificate
\end{enumerate}
\item is the service a Single Onion Service\footnote{A Single Onion Service is an Onion Service that is configured for services that do not require anonymity themselves, but want it for clients connecting to their service. Single Onion Services use only three hops in the circuit rather than the typical six hops for onion services.}
\item Strict Transport Security\cite{rfc6797}
\item Content Security Policy\cite{Sartori:23:CSP}
\item Permissions-Policy\cite{w3c-pp}
\item X-Frame-Options
\item X-Content-Type-Options
\item Referrer Policy\cite{Eisinger:17:RP}
\item Cross Origin Embedder Policy\cite{corp-coep}
\item Cross Origin Opener Policy\cite{corp-coep}
\item Cross Origin Resource Policy\cite{corp-coep}
\item Onion-Location\footnote{\url{https://community.torproject.org/onion-services/advanced/onion-location/}}
\item Apache mod\_{}status\footnote{\url{https://httpd.apache.org/docs/2.4/mod/mod_status.html}} \cite{onionscan-what}
\end{enumerate}

A tool called OnionSec was developed to complete this analysis. The source code for this tool is available on GitHub\footnote{\url{https://github.com/theenbyperor/onion-sec}}. The tool is also available as a service on the web for anyone wishing to run the same set of tests against their Onion Service and receive a report on its security\footnote{\url{https://onionsec.net}}.

\subsection{Collection of data about clear web sites}
For comparison against more standard clear web sites data collected by Scott Helme in his Crawler.Ninja\footnote{\url{https://crawler.ninja}} were used. This dataset contains security metrics of the top 1 million sites on the internet, as ranked by Amazon's Alexa.

\section{Results}
The data collected about Onion Services is available on GitHub\footnote{\url{https://gist.github.com/TheEnbyperor/90ba14517a0a8d184f8744252c6a6e8e}}. Throughout this paper a confidence of 5\% is used to reject $H_0$.

Of the 957 Onion Services discovered, 309 (32.29\%) could not be scanned for one reason or another, such as the site being offline, requiring client authentication, or taking too long to respond (over 30 seconds). Of those scanned the open ports are displayed in table \ref{table:open-ports}. The 

\begin{table}[h!]
\centering
\begin{tabular}{|c|c|c|}
\hline 
Port & Number of sites & Proportion of sites scanned \\ 
\hline 
HTTP (80) & 614 & 94.75\% \\ 
\hline 
HTTPS (443) & 198 & 30.56\% \\ 
\hline 
SSH (22) & 4 & 0.62\% \\ 
\hline 
FTP (21) & 0 & 0.00\% \\ 
\hline 
SMTP (25) & 4 & 0.62\% \\ 
\hline 
\end{tabular} 
\caption{Open ports on Onion Services}
\label{table:open-ports}
\end{table}

\paragraph{Apache \texttt{mod\_{}status}} The proportion of Onion Services with exposed Apache \texttt{mod\_{}status} pages is shown in table \\ref{table:mod-status}. This is comparable to the 7-10\% detected by OnionScan in 2017\cite{onionscan}, showing no improvement, and perhaps even a worsening on this front. No detailed data was released by OnionScan so a statistical test of this unfortunately cannot be conducted.

\begin{table}[h!]
\centering
\begin{tabular}{|c|c|c|}
\hline 
Port & Number of sites & Proportion of sites scanned \\ 
\hline 
\texttt{mod\_{}status} Server Info & 61 & 9.41\% \\ 
\hline 
\texttt{mod\_{}status} Server Status & 69 & 10.65\% \\ 
\hline 
\end{tabular} 
\caption{Apache \texttt{mod\_{}status} on Onion Services}
\label{table:mod-status}
\end{table}

\subsection{HTTPS support}

\paragraph{Hypothesis 1} Onion Services are as likely to use HTTPS as other sites. The proportion of sites supporting HTTPS are shown in \ref{table:https}. The $\chi^2$ contingent test between Onion services and other sites has a p-value of $<0.05$ ($8.16 \times 10^{-225}$), thus rejecting the null hypothesis and confirming that Onion Services are less likely to use HTTPS than sites on the clear web.

\begin{table}[h!]
\centering
\begin{tabular}{|c|c|c|c|}
\hline 
HTTPS supported & Type of site & Number & Proportion of sites scanned \\ 
\hline 
Yes & Onion Service & 198 & 30.56\% \\ 
\hline 
No & Onion Service & 450 & 69.44\% \\ 
\hline 
Yes & Clear web & 804,999 & 80.50\% \\ 
\hline 
No & Clear web & 195,001 & 19.50\% \\ 
\hline 
\end{tabular} 
\caption{HTTPS support on different sites}
\label{table:https}
\end{table}

\paragraph{Hypothesis 2} Single Onion Services are as likely to use HTTPS as other Onion Services. The proportion of sites being a Single Onion Service is shown in table \ref{table:single-onion}. The $\chi^2$ contingent test between Single Onion Services and HTTPS has a p-value of $<0.05$ ($0.00407$), thus rejecting the null hypothesis and confirming that Single Onion Sites are more likely to use HTTPS than other Onion Services.

\begin{table}[h!]
\centering
\begin{tabular}{|c|c|c|}
\hline 
Single Onion status & Number of sites & Proportion of sites scanned \\ 
\hline 
Not a Single Onion Service & 504 & 77.78\% \\ 
\hline 
Single Onion Service & 144 & 22.22\% \\ 
\hline 
Single Onion Service with HTTPS & 124 & 19.14\% \\
\hline 
Single Onion Service without HTTPS & 20 & 3.09\% \\
\hline 
Not a Single Onion Service with HTTPS & 74 & 11.42\% \\
\hline 
Not a Single Onion Service without HTTPS & 430 & 66.36\% \\
\hline 
\end{tabular} 
\caption{Single Onion Status of Onion Services}
\label{table:single-onion}
\end{table}

\paragraph{Hypothesis 3} Onion Services use the same versions of TLS as other sites. Of the Hiddden Services supporting HTTPS the proportions of them supporting different TLS versions are shown in table \ref{table:tls-version}. The same metrics for clear web sites are shown in table \ref{table:tls-version-clear}. The $\chi^2$ contingent test between Onion Services and the clear web has a p-value of $<0.05$ ($6.42 \times 10^{-35}$), thus rejecting the null hypothesis and confirming that Onion Services use different versions of TLS than on the clear web. Onion Services tend to use TLS 1.2 whilst those on the clear web have a clear preference for TLS 1.3.

\begin{table}[h!]
\centering
\begin{tabular}{|c|c|c|}
\hline 
Highest TLS version supported & Number of sites & Proportion of sites supporting HTTPS \\ 
\hline
SSL 2.0 & 0 & 0.00\% \\
\hline
SSL 3.0 & 0 & 0.00\% \\
\hline 
TLS 1.0 & 0 & 0.00\% \\
\hline
TLS 1.1 & 0 & 0.00\% \\
\hline
TLS 1.2 & 121 & 61.11\% \\
\hline
TLS 1.3 & 75 & 37.88\% \\
\hline
\end{tabular} 
\caption{TLS versions supported on Onion Services}
\label{table:tls-version}
\end{table}

\begin{table}[h!]
\centering
\begin{tabular}{|c|c|c|}
\hline 
Highest TLS version supported & Number of sites & Proportion of sites supporting HTTPS \\ 
\hline
SSL 2.0 & 0 & 0.00\% \\
\hline
SSL 3.0 & 0 & 0.00\% \\
\hline 
TLS 1.0 & 342 & 11.62\% \\
\hline
TLS 1.1 & 0 & 0.00\% \\
\hline
TLS 1.2 & 148,207 & 23.64\% \\
\hline
TLS 1.3 & 478,324 & 76.30\% \\
\hline
\end{tabular} 
\caption{TLS versions supported on clear web sites}
\label{table:tls-version-clear}
\end{table}

\paragraph{Hypothesis 4} Onion Services use the certificate key lengths as other sites. Of the Onion Services supporting HTTPS the proportions of them having different key lengths are shown in table \ref{table:key-length}. The same metrics for clear web sites are shown in table \ref{table:key-length-clear}. Only RSA public keys are considered, those not counted in the tables used Elliptic Curve or another kind of key, for which key length has less of an effect on security. The $\chi^2$ contingent test between Onion Services and the clear web has a p-value of $<0.05$ ($5.29 \times 10^{-17}$), thus rejecting the null hypothesis and confirming that Onion Services use different key lengths to the clear web. Onion Services tend to use longer key lengths than on the clear web, increasing security.

\begin{table}[h!]
\centering
\begin{tabular}{|c|c|c|}
\hline 
Key Length & Number of sites & Proportion of sites supporting RSA \\ 
\hline
< 2048 & 0 & 0.00\% \\
\hline
2048 & 52 & 62.62\% \\
\hline 
3072 & 0 & 0.00\% \\
\hline
4096 & 31 & 37.35\% \\
\hline
8192 & 0 & 0.00\% \\
\hline
\end{tabular} 
\caption{Key Length of Onion Services}
\label{table:key-length}
\end{table}

\begin{table}[h!]
\centering
\begin{tabular}{|c|c|c|}
\hline 
Key Length & Number of sites & Proportion of sites supporting RSA \\ 
\hline
< 2048 & 117 & 0.02\% \\
\hline
2048 & 445,237 & 90.39\% \\
\hline 
3072 & 3,167 & 0.64\% \\
\hline
4096 & 44,059 & 8.94\% \\
\hline
8192 & 19 & 0.00\% \\
\hline
\end{tabular} 
\caption{Key Length of clear web sites}
\label{table:key-length-clear}
\end{table}

\paragraph{Hypothesis 5} Onion Services are as likely to use Extended Validation certificates as other sites. The proportion of sites having EV certificates are shown in \ref{table:ev}. The $\chi^2$ contingent test between Onion Services and other sites has a p-value of $<0.05$ ($3.98 \times 10^{-33}$), thus rejecting the null hypothesis and confirming that Onion Services are more likely to use EV certificates than other sites.

\begin{table}[h!]
\centering
\begin{tabular}{|c|c|c|c|}
\hline 
EV certificate & Type of site & Number & Proportion of sites scanned \\ 
\hline 
Yes & Onion Service & 18 & 9.09\% \\ 
\hline 
No & Onion Service & 180 & 90.91\% \\ 
\hline 
Yes & Clear web & 7,069 & 0.88\% \\ 
\hline 
No & Clear web & 797,930 & 99.12\% \\ 
\hline 
\end{tabular} 
\caption{Extended Validation support on different sites}
\label{table:ev}
\end{table}

\subsubsection{Other data}
This section consists of data that has no comparable equivalent in the Crawler.Ninja dataset, but is nonetheless interesting and included in table \ref{table:tls-other} for reference.

\begin{table}[h!]
\centering
\begin{tabular}{|c|c|c|c|}
\hline 
Metric & Number & Proportion of sites scanned \\ 
\hline 
Supporting TLS compression & 0 & 0.00\% \\ 
\hline 
Supporting TLS 1.3 early data & 9 & 4.55\% \\ 
\hline 
Supporting TLS downgrade prevention & 196 & 98.99\% \\ 
\hline 
Vulnerable to heartbleed & 0 & 0.00\% \\ 
\hline 
Vulnerable to CCS injection & 0 & 0.00\% \\ 
\hline 
Vulnerable to client renegotiation DOS & 0 & 0.00\% \\ 
\hline 
Vulnerable to ROBOT (weak oracle) & 0 & 0.00\% \\ 
\hline 
Vulnerable to ROBOT (weak oracle) & 0 & 0.00\% \\ 
\hline 
\end{tabular} 
\caption{Other metrics concerning Onion Service's TLS setup}
\label{table:tls-other}
\end{table}

\subsection{HTTP Headers}

\paragraph{Hypothesis 6} Onion Services are as likely to use Strict Transport Security as other sites. The proportion of sites having an STS header are shown in \ref{table:sts}. The $\chi^2$ contingent test between Onion Services and other sites has a p-value of $<0.05$ ($5.00 \times 10^{-8}$), thus rejecting the null hypothesis and confirming that Onion Services are less likely to use STS than other sites.

\begin{table}[h!]
\centering
\begin{tabular}{|c|c|c|c|}
\hline 
STS Header & Type of site & Number & Proportion of sites scanned \\ 
\hline 
Yes & Onion Service & 79 & 12.87\% \\ 
\hline 
No & Onion Service & 535 & 87.13\% \\ 
\hline 
Yes & Clear web & 220,742 & 22.07\% \\ 
\hline 
No & Clear web & 779,258 & 77.93\% \\ 
\hline 
\end{tabular} 
\caption{Strict Transport Security support on different sites}
\label{table:sts}
\end{table}

\paragraph{Hypothesis 7} Onion Services are as likely to use a Content Security Policy as other sites. The proportion of sites having a CSP header are shown in \ref{table:csp}. The $\chi^2$ contingent test between Onion Services and other sites has a p-value of $<0.05$ ($6.88 \times 10^{-38}$), thus rejecting the null hypothesis and confirming that Onion Services are more likely to use CSP than other sites.

\begin{table}[h!]
\centering
\begin{tabular}{|c|c|c|c|}
\hline 
CSP Header & Type of site & Number & Proportion of sites scanned \\ 
\hline 
Yes & Onion Service & 160 & 26.06\% \\ 
\hline 
No & Onion Service & 454 & 73.94\% \\ 
\hline 
Yes & Clear web & 102,296 & 10.23\% \\ 
\hline 
No & Clear web & 897,704 & 89.77\% \\ 
\hline 
\end{tabular} 
\caption{Content Security Policy support on different sites}
\label{table:csp}
\end{table}

\paragraph{Hypothesis 8} Onion Services are as likely to use a Permissions Policy as other sites. The proportion of sites having a PP header are shown in \ref{table:pp}. The $\chi^2$ contingent test between Onion Services and other sites has a p-value of $<0.05$ ($0$ - less that the smallest floating point number in python), thus rejecting the null hypothesis and confirming that Onion Services are more likely to use PP than other sites.

\begin{table}[h!]
\centering
\begin{tabular}{|c|c|c|c|}
\hline 
PP Header & Type of site & Number & Proportion of sites scanned \\ 
\hline 
Yes & Onion Service & 116 & 18.89\% \\ 
\hline 
No & Onion Service & 498 & 81.11\% \\ 
\hline 
Yes & Clear web & 5,623 & 0.56\% \\ 
\hline 
No & Clear web & 994,377 & 99.44\% \\ 
\hline 
\end{tabular} 
\caption{Permissions Policy support on different sites}
\label{table:pp}
\end{table}

\paragraph{Hypothesis 9} Onion Services are as likely to use X-Frame-Options as other sites. The proportion of sites having an XFO header are shown in \ref{table:xfo}. The $\chi^2$ contingent test between Onion Services and other sites has a p-value of $<0.05$ ($8.44 \times 10^{-42}$), thus rejecting the null hypothesis and confirming that Onion Services are more likely to use XFO than other sites.

\begin{table}[h!]
\centering
\begin{tabular}{|c|c|c|c|}
\hline 
XFO Header & Type of site & Number & Proportion of sites scanned \\ 
\hline 
Yes & Onion Service & 305 & 49.67\% \\ 
\hline 
No & Onion Service & 309 & 50.33\% \\
\hline 
Yes & Clear web & 256,952 & 25.70\% \\ 
\hline 
No & Clear web & 743,048 & 74.30\% \\ 
\hline 
\end{tabular} 
\caption{X-Frame-Options support on different sites}
\label{table:xfo}
\end{table}

\paragraph{Hypothesis 10} Onion Services are as likely to use X-Content-Type-Options as other sites. The proportion of sites having an XCTO header are shown in \ref{table:xcto}. The $\chi^2$ contingent test between Onion Services and other sites has a p-value of $<0.05$ ($1.61 \times 10^{-123}$), thus rejecting the null hypothesis and confirming that Onion Services are more likely to use XCTO than other sites.

\begin{table}[h!]
\centering
\begin{tabular}{|c|c|c|c|}
\hline 
XCTO Header & Type of site & Number & Proportion of sites scanned \\ 
\hline 
Yes & Onion Service & 374 & 60.91\% \\ 
\hline 
No & Onion Service & 240 & 39.09\% \\
\hline 
Yes & Clear web & 215,687 & 21.57\% \\ 
\hline 
No & Clear web & 784,313 & 78.43\% \\ 
\hline 
\end{tabular} 
\caption{X-Content-Type-Options support on different sites}
\label{table:xcto}
\end{table}

\paragraph{Hypothesis 11} Onion Services are as likely to use a Referrer Policy as other sites. The proportion of sites having an RP header are shown in \ref{table:rp}. The $\chi^2$ contingent test between Onion Services and other sites has a p-value of $<0.05$ ($4.53 \times 10^{-22}$), thus rejecting the null hypothesis and confirming that Onion Services are less likely to use an RP than other sites.

\begin{table}[h!]
\centering
\begin{tabular}{|c|c|c|c|}
\hline 
RP Header & Type of site & Number & Proportion of sites scanned \\ 
\hline 
Yes & Onion Service & 0 & 0.00\% \\ 
\hline 
No & Onion Service & 614 & 100.00\% \\
\hline 
Yes & Clear web & 133,314 & 13.33\% \\ 
\hline 
No & Clear web & 866,686 & 86.67\% \\ 
\hline 
\end{tabular} 
\caption{Referrer Policy support on different sites}
\label{table:rp}
\end{table}

\subsubsection{Other data}
This section consists of data that has no comparable equivalent in the Crawler.Ninja dataset, but is nonetheless interesting and included in table \ref{table:http-other} for reference.

\begin{table}[h!]
\centering
\begin{tabular}{|c|c|c|c|}
\hline 
Metric & Number & Proportion of sites scanned \\ 
\hline 
Cross Origin Embedder Policy present & 46 & 7.49\% \\
\hline 
Cross Origin Opener Policy present & 50 & 8.14\% \\
\hline 
Cross Origin Resource Policy present & 52 & 8.47\% \\
\hline 
Onion-Location present & 111 & 18.08\% \\
\hline 
\end{tabular} 
\caption{Other metrics concerning Onion Service's HTTP setup}
\label{table:http-other}
\end{table}

\section{Discussion}
Tor Onion Services are primarily used for hosting HTTP based web sites, but some are used for other purposes, thus the security of how HTTP is configured on these Onion services is paramount to the security of their users. 

Onion Services are not configured in any metric in a similar way to sites on the clear web. Onion Services are less likely to support HTTPS, and when they do support it are likely to not support never versions of the TLS protocol. In spite of this Onion Services do tend to use longer RSA key lengths than sites on the clear web, improving the security of the connection. They're also more likely to use Extended Validation Certificates, providing greater assurances to the service's user about who they're talking to, which important in cases such as whistle-blowing to news organisations.

Considering application layer security defences in HTTP Onion Services are more likely to use all but two security policies - Strict Transport Security and Referrer Policy. The reason for the complete lack of Referrer Policies on Onion Services is not clear, however the lack of STS can be explained by the general lack of HTTPS on Onion Services which is a prerequisite for enabling STS.

Overall Onion Services do more to protect their users than sites on the clear web, however are lacking in HTTPS deployment. This can be explained by the lack of Certificate Authorities that issue certificates for Onion Services, and that none offer certificates free of charge\footnote{DigiCert and HARICA are the only CAs to issue such certificates. DigiCert additionally only issues the much more expensive EV certificates. None of the common free CAs such as Let's Encrypt issue such certificates}. More work needs to be done to increase the level of HTTPS deployment on Onion Services to a similar level as on the clear web.

In terms of what could be explored in future work; this research did not look at security vulnerabilities in Onion Services such as Cross Site Scripting attacks, SQL injection, or other such vulnerabilities which are errors of poor coding than of security policy. Additionally it was raised by some (after data collection) that some less knowledgeable users may publish their sites on ports such as 5000, 8080 and 8443. Finally, no tests of the availability of TLS/STARTTLS on SMTP nor FTP where conducted.

\appendix{}

\section{An onion}
\includegraphics[scale=0.5]{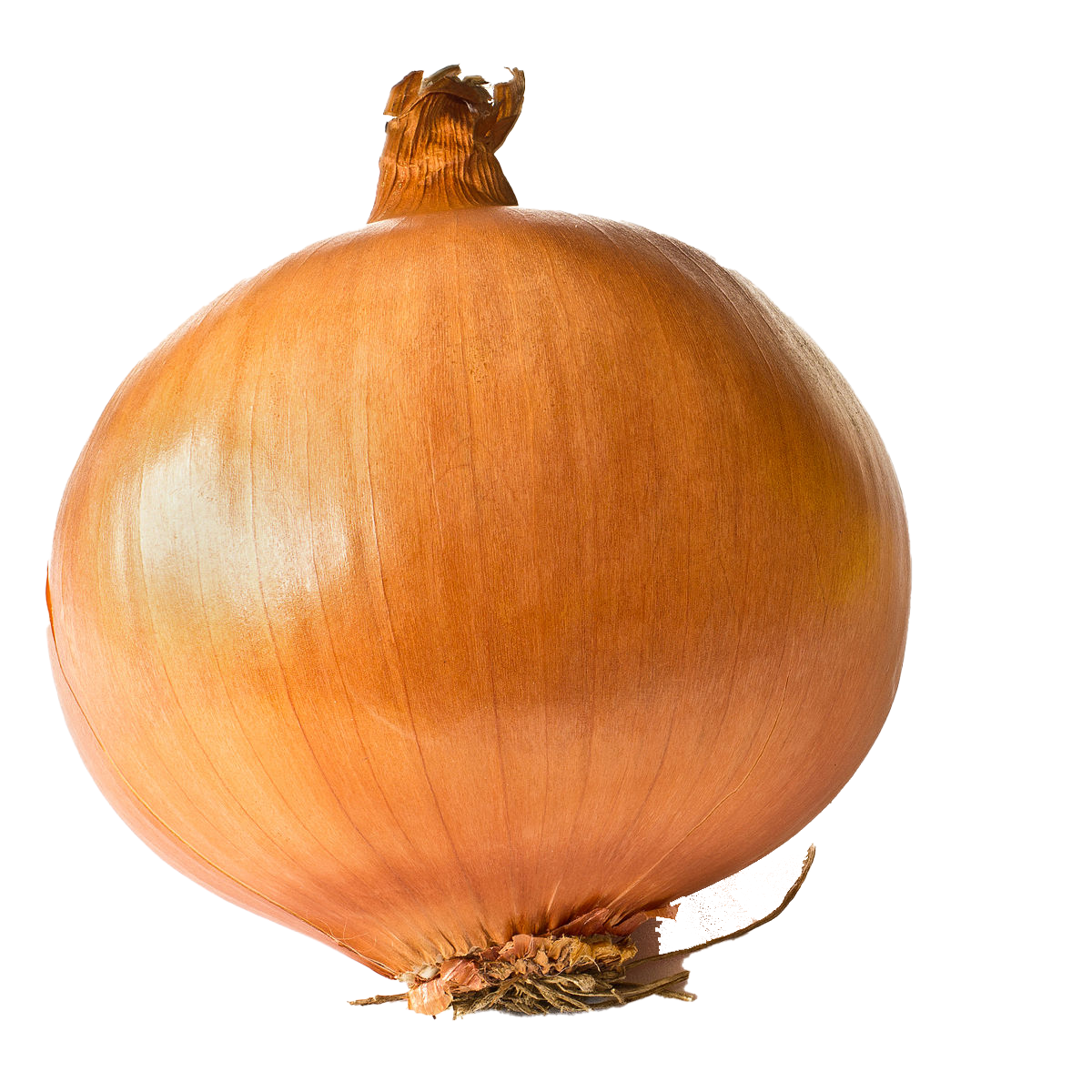}

\section{Acknowledgements}
With thanks to Iain Learmonth\footnote{\url{https://irl.xyz}} for providing guidance on the methodology for collecting a list of Onion Services, and all those at the Max-Planck-Institut für Informatik\footnote{\url{https://www.mpi-inf.mpg.de/}} who provided feedback on an earlier revision of this work presented there. 

\printbibliography{}

\end{document}